\documentclass{article}

\usepackage[preprint]{neurips_2026} 

\usepackage[utf8]{inputenc} 
\usepackage[T1]{fontenc}    
\usepackage{hyperref}       
\usepackage{url}            
\usepackage{booktabs}       
\usepackage{amsfonts}      
\usepackage{nicefrac}        
\usepackage{microtype}      
\usepackage{xcolor}         

\usepackage{graphicx}
\usepackage[most]{tcolorbox}
\usepackage{amssymb} 
\usepackage{algorithm}
\usepackage{algorithmic}
\usepackage{amsmath}

\usepackage{colortbl}

\usepackage{caption}
\usepackage{multirow}
\usepackage[T1]{fontenc} 

\def\BibTeX{{\rm B\kern-.05em{\sc i\kern-.025em b}\kern-.08em
    T\kern-.1667em\lower.7ex\hbox{E}\kern-.125emX}}

\newcommand \ignore[1]{}

\title{Stop using Conventional Metrics to Evaluate
Membership Inference Attacks on LLMs: A Call for
Holistic Benchmarks}

\author{
Clyde Baidoo \\
\texttt{clbai4@morgan.edu}
\And
Sayanton Diboo \\
\texttt{sdibbo@ua.edu}
}

\begin{document}

\maketitle

\begin{abstract}
  When membership inference attacks (MIA) used on large language models (LLMs) are unavailing, they introduce extreme risks of model vulnerability and privacy leakage. These risks are amplified particularly when an ineffective MIA model is not captured. In larger distributional scales, such models are reported to fail incoherently when they are deployed, even though they demonstrate good performance evaluations post-training. We term this inconsistency as “silent failures”. The implication is that the use of traditional evaluation metrics creates misleading impressions of the vulnerability and robustness of MIAs in LLMs.

\textbf{This position paper argues that traditional evaluation metrics including accuracy and Area Under the Curve (AUC) are fundamentally insufficient for assessing membership inference attacks on large language models, and that the field must adopt a standardized, architecture-aware holistic benchmark.} We discuss how recent studies rely on high accuracy to depict the reliability, safety, privacy-oriented, and robustness of a model, yet they frequently fail to reflect true memorization signals when deployed. Our careful deductions help to bridge the gap between a model's reflection of good performance numbers and real-world model stability. This enables studies to ensure a more accurate representation and evaluation of the reliability of MIAs in LLMs.

\end{abstract}

\section{Introduction}
Membership inference attacks help to evaluate privacy leakage in machine learning. They query a machine learning model to detect whether a specific input was part of the model's training data~\cite{shokri2017membership,yeom2018privacy}. 
Recent studies rely on their facts on the performance metrics of membership inference attacks, mostly on traditional metrics such as accuracy and AUC. However, when such traditional metrics are used to evaluate the performance of an MIA model, they do not reflect the true memorization risk or privacy leakage of large language models. This becomes a major concern for LLMs, and we address this by answering the meta-question: \textit{are current evaluation metrics for Membership Inference Attacks on LLMs measuring true memorization, or are they producing misleading signals due to high performance indicators and specific metric choice, scale effects, and evaluation artifacts?} We examine four representative studies and systematically analyze their methodologies and the underlying assumptions and pitfalls. 
The success of MIA from these studies on LLMs mainly depends on the attack performance metrics, which persistently fail to capture the real-world vulnerability risks of LLMs. Additionally, a significant number of studies adopt the construction of a post-hoc dataset, architectural differences, and scaling effects that also depict mild inflation on the perceived nature of attack performance without demonstrating significant leakages~\cite{song2017machine, niu2025comparing}.

In addressing this, our focus shifts from identifying and recommending stronger attack mechanisms to analyzing the evaluation processes or methodologies, which helps us to pinpoint specific metrics that are more susceptible to silent failures in real-world contexts and under what conditions they could fail. The four representative studies considered for our analysis include~\cite{liao2025perprob, puerto2025scaling, hayes2025exploring, duan2024membership}, which make use of a variety of threat models, methodologies, and LLM scales. 

By structuring our analysis across diverse approaches, we are able to clarify why over-reliance on traditional metric evaluation alone is insufficient. Relying on traditional evaluation metrics could mask nuances by reflecting dataset biases and being influenced by the model architecture rather than true memorization signals.  \textbf{We argue that current evaluation metrics for Membership Inference Attacks on LLMs systematically fail to measure true memorization risk, and that the community must adopt a standardized holistic benchmark that is architecture-aware, multi-metric, and resistant to dataset artifacts.} This subsequently enables us to reveal commonly overlooked pitfalls in recent MIA evaluations. We also spotlight the limitations of pre-existing benchmarks, and we propose the development of a more robust, context-aware, and architecturally relevant deployment framework. Our deductions help in understanding why MIAs in modern LLMs appear superficially robust due to strong metric performance but fail to showcase meaningful memorization risk in real-world scenarios.

\section{Background}
Machine Learning has been rapidly advancing recently, especially with the leverage provided by Generative Pretrained Transfomers (GPT)~\cite{radford2018improving, yenduri2024gpt}. Most of these rapid advances require large-scale datasets for training of ML models. The content of the training data spans from regular data to sensitive information, including controlled unclassified information (CUI), health records, financial data, and sensitive personal information, which are usually only used for official or internal purposes. Models must be trained on large-scale data to optimize generalization and also capture patterns and dimensions that improve memorization. However, these data become primary subjects to external influence, including adversarial attacks, backdoor attacks, model evasion, and poisoning attacks. In this position paper, our main focus is on membership inference attacks, where an attacker's motive is to determine whether specific data were part of the model's training data. Although our review focuses on four main representative studies, Liao et al.~\cite{liao2021we} provide concrete evidences that evaluation bottlenecks and silent failures have been recurring across major domains in language processing, vision and recommender systems, and reinforcement learning. 

\subsection{Membership Inference Attack}
\begin{figure*}[t] 
\centering \includegraphics[width = \textwidth, keepaspectratio]{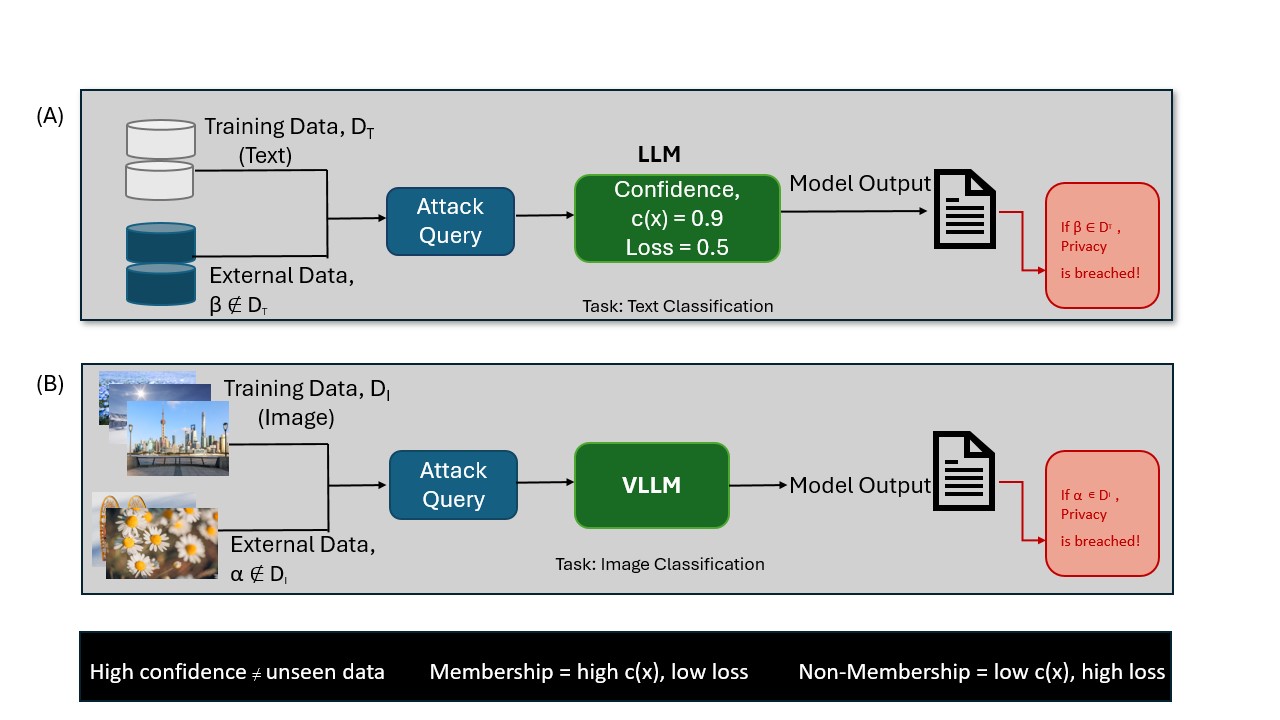} \caption{Membership Inference Attacks in (A) Texts-based LLMs   (B) Image-based LLMs}
\label{fig:MIA} 
\end{figure*}

Membership Inference Attacks exploit the memorization abilities of machine learning models to distinguish between member and non-member data samples within the training 
data~\cite{shokri2017membership, yeom2018privacy, carlini2022membership}. A typical example of an MIA involves the unauthorized use of person's health records that are obtained from ML models trained for clinical purposes~\cite{wu2025membership}. When an attacker accesses data that is deemed sensitive or confidential, they misuse them for illegal schemes. Additionally, when predictive models are trained on open source codebases that are injected with code poisoning attacks, the model becomes a target for MIAs and adversarial attacks~\cite{puerto2025scaling, li2025advancing}.
In Figure~\ref{fig:MIA}, we present the overview of MIA attacks in LLMs based on images and text.

\noindent If \textbf{B} is a binary classifier and \begin{equation}
    k = \{q, r\}
\end{equation}

\noindent Given a model \textbf{M} and a target data sample \textbf{k}, \textbf{q} denotes input features and \textbf{r} represents the label, an MIA predicts member (1) and non-member (0) as
\begin{equation}
    B(M, k) = \{0, 1\}
\end{equation}

\begin{equation}
    {B}(c, k) =
    \begin{cases}
        1 & \text{if } c(x) \ge \tau \\
        0 & \text{otherwise}
    \end{cases}
\end{equation}

\noindent where $c(x)$ is the confidence of a model in its prediction and $\tau$ is a threshold.

\begin{equation}
    c(x) \ge \tau 
\end{equation}
MIAs use this as an edge to exploit membership data within the training set, which violates confidentiality and privacy regulations~\cite{niu2024survey,10.1145/3523273}. 

\subsection{Memorization vs Generalization}
In memorization, deep neural networks (DNNs) learn specific details or particular features of examples~\cite{wei2025memorization}. Memorization serves as a useful indicator for MIAs as it only attributes greater confidence to specific rare training samples~\cite{carlini2022quantifying}. For autoregressive models, 
\begin{equation}
    \log Q_{\theta}(x) = \sum_{m=1}^M \log Q_{\theta}(x_m \mid x_{<m})
\end{equation}
where $log Q_{\theta}(x)$ is the total log-likelihood of the sequence with $\theta$ as parameter, $x$ represents the entire sequence, $x_m$, the $m^{th}$ token in the sequence, and $x_{<m}$ represents the prefix notations preceding m.  
Carlini et al.~\cite{carlini2022quantifying} introduced $Exposure$, which also helps to capture tail memorization. 
\begin{equation}
    Exposure(x) = \log_2 \mid R \mid - \log_2 rank_\theta (x)
\end{equation}
where $|R|$ is the randomness of space, $rank_\theta (x)$ is the rank when the model makes a prediction $\theta$.

Generalization, on the other hand, measures how well models perform on trained sample data.
Given 
\begin{equation}
    k =(q, r) \backsim P,loss \ l(f, k) \epsilon [0, 1] 
\end{equation}
\begin{equation}
    \hat{R}(f) = \frac{1}{n} \sum_{i=1}^{n} \ell(f, k_i)
\end{equation}
\begin{equation}
    R(f) = \mathbb{E}_{k \sim P} \big[ \ell(f, k) \big]
\end{equation}
\begin{equation}
    \Delta_{\text{gen}} = R(f) - \hat{R}(f)
\end{equation}
where $l(f, k) \epsilon [0, 1]$ is a loss function, $\hat{R}(f)$ is the Empirical risk, $R(f)$ is the true or expected risk, and $P$ is a probability distribution.

A large generalization gap, $\Delta_{\text{gen}}$, indicates overfitting and a higher risk of MIA vulnerability. In addition, generalization errors scale the MIA advantage, essentially increasing privacy leakages~\cite{shokri2017membership, carlini2022membership}. This implies that, although models can perform well in generalization, they cannot capture memorization in rare sequences. Duan et al.~\cite{duan2024membership} proved that MIAs generally perform weakly on large LLMs, yet they are still able to achieve greater success on targeted samples. Memorization is necessary, as it creates tail privacy leakages even when the evaluation metrics appear strong. However, it becomes malignant when the data contain rare sensitive strings, and weak training filtering~\cite{carlini2022quantifying}. This accounts for silent failures when they are deployed after evaluation, even when
\begin{equation}
    \Delta{\text{gen}} \approx 0
\end{equation}
If $f\theta$ is a trained model, $l(f, k)$ a bounded loss, $L_{member}$ and $L_{non-member}$ define random loss variables with $S$ samples, then silent failures can be depicted as:
\begin{equation}
    L_{member} = l(f, k), k \epsilon S
\end{equation}
\begin{equation}
    L_{non-member} = l(f, k), k \backsim P
\end{equation}
\begin{equation}
    silent failure = \Delta_{mean}small , \ \Delta_{tail}large
\end{equation}
\begin{equation}
    Let: \\
    \ \Delta_{mean} = \mathbb{E}{[L_{non-member}] - \mathbb{E}{[L_{member}]}}
\end{equation}
\begin{equation}
    \Delta_{mean} = R(f) - \hat{R}(f)
\end{equation}
\begin{equation}
    \Delta_{tail}(q) = Q_{q} (L_{non-member}) - Q_{q}(L_{member})
\end{equation}
where $Q_q$ is the level of the loss distribution $q-quantile$. Tail memorization occurs when $\Delta_{tail}(q)$ >> $\Delta_{mean}$~\cite{carlini2022quantifying}. \\
Evaluating only the mean case of MIA performance would fail to fully reflect the real risk of privacy leakage in real-world deployment settings. When generalization and memorization are accurately accounted for in MIAs, they help to control the average and worst-case vulnerabilities, respectively.

\section{Methodology}
We conduct a systematic review of recent research works geared towards MIA failures and post-deployment analysis, unforeseen architectural and technical debts in machine learning systems, model degradation, and underspecification of robustness in model leakages and inferences. In our analysis, we considered 50 research publications that closely aligned with our inclusion criteria and established a solid claim of model failures after training. We performed a comparative analysis on 4 main representative studies that explored traditional evaluation metrics for MIA models and analyzed their correlational results.

\subsection{Inclusion Criteria}
In selecting research papers to be analyzed, our main consideration was limited to only recent works that emphasized traditional metrics for evaluating MIAs. In essence, our core claim of model failure post deployment was established by reviewing 107 research publications in addition to an extensive review of the revelations established by Liao et al.~\cite{liao2021we}. Our review was narrowed to only four representative studies. Our main interest was in research papers that employed traditional metrics for their evaluation analysis.  We also narrowed our search to papers that had core contributions in the fundamental dimensions and areas that directly affect MIA performances.

\subsection{Dataset}
Although our representative studies were reduced to 4 main representative studies, our inclusion criteria allowed us to review a respectable number of 50 additional research publications. Most of these research articles were published between 2023 and 2025, with a few others in 2017 through 2019.

Across all representative studies, one noticeable recurring pattern across the datasets used was the collection of post-hoc datasets, particularly in black-box settings. We found this specific integration to be concerning particularly in events when the member and non-member sets are not distributionally matching. This is because MIAs exploit such distributional shift differences instead of memorizing specific training examples\cite{even2026causal}. An additional number of exploits could span spurious signals usually unrelated to memorizations, including token differences, topic bias, and artifacts. When there is a significant difference in the distributional match of member and non-member sets, it inflates traditional metrics such as accuracy and AUC instead of true memorization of specific training examples. This implies that distribution changes can also influence higher performance metrics, but not a true reflection of higher memorization~\cite{shokri2017membership, hayes2025exploring, duan2024membership}. The effect inflates the accuracy and AUC even when the model exhibits absent or minimal memorization~\cite{salem2018ml, truex2019demystifying}.

Duan et al.~\cite{duan2024membership} also aligned member and non-member distributions across the same domain.  Their results indicate a significant decrease in MIA performance and attack advantage.

\subsection{Evaluation Dimension}
The core claim of this position paper is that traditional evaluation metrics are insufficient for analyzing membership inference attacks on large language models and that relying solely on good performance numbers during evaluation for the deployment of an MIA model should be deemed as an unreliable practice.  We support our claims by analyzing a number of key dimensions that are substantial to the validity of MIA results and significantly affect their explainability. These dimensions are sectioned into attack type, model architecture, dataset setup, and evaluation metrics. 
Our first consideration is the type of attack. This helps us to perform a detailed analysis of the strength of the model (e.g., black-box, white-box). We also investigated the model’s architecture to track the memorization dynamics (encoder-only vs. decoder-only) and how the dataset is structured. Analyzing the dataset setup helps us to avoid oddly structured setups which usually influence true memorization signals. It also prevents  the inclusion of irrelevant artifacts that could influence MIA results.
In addition, we also focus on metric-centered evaluations, including accuracy, AUC, True Positive Rates (TPR), False Positive Rates (FPR), attack advantage, and perplexity. Our choice of metrics is inferred directly from the representative studies, and they enable us to capture the myriad aspects of model inference and failure. Finally, we explore the baseline risks and analyze the scale of the model and the size of the dataset. The model and data scaling help us to examine the true efficiency of various attack mechanisms and their subsequent effects on generalization and overfitting. The baseline risks also help us define a successful MIA attack by identifying bias and dataset leakage.

\section{Comparative Analysis of MIAs on LLMs}

\begin{table*}[t]
\centering
\caption{Comparison of key MIA studies on LLMs across core dimensions.}
\label{tab:mia_comparison}
\scriptsize
\begin{tabular}{|p{1.8cm}|p{1.5cm}|p{1.5cm}|p{1.3cm}|p{1.5cm}|p{3cm}|}
\hline
\textbf{Paper} & \textbf{Attack Type} & \textbf{Model Architecture} & \textbf{Dataset Split} & \textbf{Metrics Used} & \textbf{Key Notes / Limitations} \\
\hline
PerProb~\cite{liao2025perprob} & 
Black-box/ perplexity-based & 
Encoder-only (BERT-style) & 
Post hoc & 
Accuracy, AUC & 
Sensitive to prompt length; metrics reflect familiarity rather than memorization. \\
\hline
Scaling Up MIAs~\cite{puerto2025scaling} & 
Black-box + reference models & 
BERT / GPT & 
Mixed, post-hoc & 
Accuracy, AUC, Attack Advantage & 
Absolute risk near randomness. \\
\hline
Do MIAs Work?~\cite{duan2024membership} & 
Black-box & 
GPT variants & 
Controlled & 
Accuracy, AUC, TPR, FPR & 
Explicit sanity checks; demonstrates scaled weak performances. \\
\hline
Limits of Strong MIAs~\cite{hayes2025exploring} & 
White-box + reference models & 
GPT / autoregressive & 
Post-hoc & 
AUC, Attack Advantage & 
Theory-based attack assumptions; upper-bound estimates are inflated. \\
\hline
\end{tabular}
\end{table*}

In carrying out the comparative analysis that features~\cite{liao2025perprob, puerto2025scaling, hayes2025exploring, duan2024membership}, we extensively study how their respective approaches and evaluation performances capture memorization in LLMs. We extend this analysis beyond the comparison of just their raw attack performances, but also place emphasis on their architectural setup, dataset construction, scalability, and metric choice sets. In Table \ref{tab:mia_comparison} and Table \ref{tab:mia_dimensions}, we show an overview of how the various dimensions are analyzed in all four representative studies.

\subsection{Cross-Paper Comparison}
In Table \ref{tab:performance_metrics} and Table \ref{tab:metric_data}, we illustrate the evaluation metrics used in the representative studies and their performance numbers in the most dominant metric used in all the studies. 
\begin{table}[h]
\centering
\caption{Metrics used across the four representative studies.$^{**}$}
\label{tab:performance_metrics}
\scriptsize
\begin{tabular}{|l|c|c|c|c|c|}
\hline
\textbf{Paper} & \textbf{Acc.} & \textbf{AUC} & \textbf{Perp.} & \textbf{AA} & \textbf{TPR, FPR} \\
\hline
PerProb~\cite{liao2025perprob} & \checkmark & \checkmark & \checkmark & \checkmark & \\
\hline
Scaling Up MIAs~\cite{puerto2025scaling} & \checkmark & \checkmark & & \checkmark & \\
\hline
Limits of Strong~\cite{hayes2025exploring} &\ & \checkmark & & \checkmark & \\
\hline
Do MIAs Work?~\cite{duan2024membership} & \checkmark & \checkmark & & &\checkmark \\
\hline
\end{tabular}
\\[2pt]
\footnotesize{**Acc.: Accuracy; Perp.: Perplexity; AA: Attack Advantage.}
\end{table}

\begin{table}[h]
\centering
\caption{Performance evaluation of the representative studies.}
\label{tab:metric_data}
\scriptsize
\begin{tabular}{|l|c|}
\hline
\textbf{Paper} &\textbf{AUC} \\
\hline
PerProb~\cite{liao2025perprob} & - \\
\hline
Scaling Up MIAs~\cite{puerto2025scaling} &\texttildelow 0.7 \\
\hline
Limits of Strong~\cite{hayes2025exploring} &$<$ 0.7 \\
\hline
Do MIAs Work?~\cite{duan2024membership}&\texttildelow 0.5 \\
\hline
\end{tabular}
\\[2pt]
\end{table}

\subsection{Metric Centric Evaluations Analysis}\label{Over-Reliance on Accuracy and AUC}
After a careful consideration of all four representative studies of MIA of LLMs, we find that the most recurring metric used to evaluate the performance is AUC.
However, the traditional metrics used in studies, including accuracy, AUC, perplexity, attack advantage, TPR, and FPR are limited to showing a complete reflection of the exact performance of MIAs.  Accuracy is highly sensitive to class imbalance, and a 50/50 member-non-member split still fails to reflect real-world threat models~\cite{shokri2017membership,salem2018ml,truex2019demystifying}. Accuracy also depends on the threshold and provides poor estimations of privacy risks, especially for LLMs that are trained on large datasets whose net results are often within excited states~\cite{hayes2025exploring, duan2024membership}.

LLMs with strong AUC scores handle false positives and false negatives in the same way, leading to high inaccuracies~\cite{yeom2018privacy,carlini2022membership}.
Previous studies show that such models fail when they are highly restricted to false positive regions while acknowledging negligible leakages regardless of considerable aggregate scores~\cite{puerto2025scaling, hayes2025strong, chen2025statistical}.
Moreover, attacks that barely outperform random guessing, particularly on scale, may still appear effective when evaluated solely using these metrics. As a result, reliance on these metrics alone makes them insufficient for evaluation.

Both accuracy and AUC fail to capture the confidence score and calibrations. Mireshghallah et al.~\cite{mireshghallah2022empirical}, Mattern et al.~\cite{mattern2023membership}, and Chen et al.~\cite{chen2025statistical} provide substantial evidence that shows MIAs weakly separate member from non-member and produce low-calibrated scores that are unreliable. This becomes particularly concerning for generative models, where the perplexity values vary across different contexts and prompts more specifically in contexts where they are reported to exclude standardization of the attack surfaces~\cite{tao2025token, chang-etal-2025-context}.
 Values therefore become inconsistent and incomparable in encoder-only, decoder-only, and vision language models, within black-box and white-box settings~\cite{wu2025membership, shejwalkar2021membership, li2024membership}. Hence, adopting a holistic evaluation benchmark specifically for MIA models would consequently improve the reliability and accuracy of the models.

\subsection{Architectural and Scaling Effects}
In Figure~\ref{fig:attack_pipeline}, we illustrate an overview of the robust attack pipeline proposed for MIAs.

The architectural and scaling effects of BERT-style models and autoregressive models should also be considered as an indispensable evaluation criterion. However, we do not recommend the practice of amalgamating both encoder-only and decoder-only models. Combining different model types introduces structural inconsistencies, which scales additional empirical risk and population risks. This is because there could be substantial variabilities in the training objectives such as the respective outputs and plausible alignments after training. Encoder-only models are usually trained with masked language models, facilitating token losses, while decoder-only models are trained with an autoregressive process, which expedites smoother likelihood surfaces. The results obtained from the architectural difference should be referenced as a proof of good MIA performance for BERT, which is an encoder-only model. However, the same results would not be reliable for autoregressive LLMs with decoder-only, such as GPT-2.

\subsection{Cross Paper Patterns}
In Table \ref{tab:mia_observations}, we highlight the key observations shown in all studies.

 \textit{MIA success is sensitive to dataset construction}: The success of an attack is dependent on how the dataset utilized is distributed, rather than how well the model truly memorizes the data. Our attention was also drawn to the actual sources of the member and non-member distributional data. If the sources are different, it becomes easier for an attack to exploit any leakage. This implies that a successful attack might still have leakages on larger scales due to time-period differences.

\textit{Metrics such as accuracy and AUC obscure absolute weakness at scale}: When the training data form a tiny fraction of the possible inputs (as with LLMs), a high AUC could correspond to an attack that produces mostly false alarms and cannot reliably identify real members.
    
\textit{Architectural differences invalidate cross-paper comparisons}: Our comparative analysis was performed across various types of architecture, which means that the results of one architectural style may not necessarily be applicable to another. For example, BERT-style models~\cite{devlin2019bertpretrainingdeepbidirectional} and GPT-style models~\cite{carlini2021extracting} possess different learning mechanisms and report different signals to the attacker. Using them interchangeably presents a flawed approach to comparison.

\textit{The lack of a standardized holistic benchmark prevents cumulative progress}: The lack of standard benchmarks slows the process of cumulative progression. All representative studies considered evidently utilized different datasets, threat models, and evaluation setups, making it difficult to make significant comparisons. Without standard benchmarks, the study keeps constantly re-discovering results instead of improving pre-existing ones.

\subsection{Implications}
The major implication inferred from the representative studies is that using traditional evaluation metrics to analyze MIA performance does not accurately capture true MIA performance. Although most recent studies accurately implement diverse methodologies, they are still unreliable, particularly in scenarios where they often fail silently after deployment in real-world settings.  Our analysis reveals concentric details on why MIA evaluations sporadically fail to measure what they truly claim to measure.
Essentially, it augments our attempt to answer our core meta quest which helps us to deduce that without an artifact-resistant dataset, architecture-aware benchmarks, and deployment-relevant metrics, performance numbers alone remain insufficient to fully assess and evaluate MIAs in LLMs.

\textit{MIA Evaluation Should Adopt Multi-Metric Reporting as a Norm, Not an Exception}: From our deductions, we emphasize that no single metric is sufficient to fully characterize membership inference risks in LLMs. Evaluations should be done with the adoption of multi-metric reporting to capture all operational dimensions and relevances on all occasions. A significant observation also indicates that while privacy risks are often dominated by extreme false-positive rates in applicable settings, only a few reports account for this. As a result, many attacks that marginally outperform random guessing appear meaningful outside of baseline contexts. We argue that MIA reports should report true positive rates at low false-positives (e.g., TPR, FPR \(\le \) 1\%). They should also consider the inclusion of calibration errors and random guessing or perplexity-only heuristics.

\subsection{Architecture-Aware Benchmarks}
Considerable difficulties and limitations were encountered in our attempts to compare MIA results across different model architectures. Recent studies have convincingly shown that MIAs success is sensitive and highly dependent on evaluation setup~\cite{puerto2025scaling, li2024membership}. This is because their systems are configured with diverse training objectives and memorization behaviors. However, multiple evaluations consider it under the single assumption of “MIA success”. If standardized benchmarks are developed that are capable of explicitly accounting for notable architectural and training-stage differences, then the comparison of MIA results becomes fundamentally plausible in the context of LLMs. Hence, we strongly encourage and propose the development of a unified privacy benchmark for LLM (e.g., extending initiatives like Privacy Bench or developing LLM-specific MIA suites) and an explicit separation of evaluation stages through pretraining, fine-tuning, or instruction tuning Retrieval-Augmented generation (RAG). Separating these stages would provide transparency and explainability of how and where privacy risk arises, which would substantially prevent over-generalization.

\section{Post-Hoc Dataset Bias}
Post hoc dataset bias refers to the construction of member and non-member sets after model training, which typically occurs due to different domain constraints and preprocessing pipelines. Recent MIA studies employ the use of unknown webscale data in model pretraining before reassembling member and non-member sets, indicating that their construction was not defined in the original model architectural setting. Essentially, it introduces distributional encoding, which results in the improvement of spurious correlations, which are targeted by attacks instead of true memorization. MIAs are heavily dependent on specific signals such as the log-likelihood gap, loss, and confidence score, which are generally vulnerable to the distributional match. This results in improved accuracy, AUC, and attack advantage in the absence of sample-level memorization. Early studies presented detailed evidence on such distributional mismatches that affect MIA attacks~\cite{shokri2017membership, puerto2025scaling, duan2024membership}. Another methodological threat, which we also observed, is the inability of accuracy, AUC, TPR, and FPR to differentiate between memorization effects and the construction of dataset artifacts. They can only be used to measure the discriminability between two sets, which affects MIA results. Stronger evaluations should incorporate a time-aligned sampling of members and non-members from the same temporal sampling window.

\section{Alternative Views}
To improve on metrics used for MIAs, we conclude by enunciating a set of position statements intended to easily accelerate future work and progress efforts to evaluating membership inference attacks for foundational models. Although we do not propose a single ``correct” evaluation metric protocol in our reflections, we shed light on identifying minimal standards and the adoption of a holistic benchmark to evaluate MIAs to avoid misleading conclusions. This would help facilitate the progression of relevant discussions that revolve around improving MIA evaluations in LLMs. However, we also acknowledge that the development of a new standard holistic benchmark that entirely differs from traditional approaches to evaluating MIAs may seem harder to adopt and may receive low acceptance by the community due to a number of considered alternative views.

\textit{Concrete Data on Established Qualitative Metrics}: We recognize the shared level of conservatism that is undoubtedly exhibited by research communities in their reluctance to adopt a novel framework, particularly in contexts when there are well-curated datasets, robust empirical evidence, and well-established evaluation standards are already in existence. For example, traditional metrics such as AUC and accuracy are well-understood and deeply ingrained in our current research ecosystem, which facilitates interpretability, transparency, and reproducibility. Adoption and a switch to a novel benchmark introduce its own reproducibility problems and also create friction in the reception of widely accepted conventional metrics.

\textit{Adoption of a Novel Benchmark:} Theoretically, the development and adoption of a standardized holistic benchmark for evaluating MIAs sounds worthwhile. Despite the theoretical advantages it asserts, developing such a comprehensive framework presents significant challenges, including difficulties in consensus-building and methodological alignment with the parties involved. \\ \\

\subsection{Alternative Holistic Benchmark}

In Algorithm~\ref{alg:halob_mia}, we introduce \textsc{HALOB-MIA} (Holistic Attack Leakage Objective Benchmark for Member Inference Attacks), an alternate holistic attack benchmark that combines multi-faceted dimensional MIA signals and feeds them to a logistic regression framework to extract relevant metric weights. We feed it into a logistic regression model to normalize the feature weights ($\alpha_i$ coefficients), ease the interpretability, and condition our holistic score to fully disintegrate into the various relevant metric contributions. This improves the transparency, explanability, and reproducibility of our results. It also unifies all the propositions in this position paper and provides an estimated weight of membership leakages across all types of attack and architectural types.

\begin{algorithm}
\caption{\textsc{HALOB-MIA}: Holistic Attack Leakage Objective Benchmark}
\label{alg:halob_mia}
\small

\begin{algorithmic}[1]
\REQUIRE Trained model $f$, member set $m$, non-member set $\bar{m}$, feature leakage functions $\{x_i(\cdot)\}_{i=1}^5$
\ENSURE Holistic leakage score, $\textsc{HALOB-MIA}$ \\

\STATE\textbf{Goal:} Quantify membership leakage of $f$ with holistic leakage score.

\STATE Construct dataset \begin{equation}
\mathcal{D} = \{(z, y)\}
\end{equation}
where
\begin{equation}
y =
\begin{cases}
1 & \text{if } z \in m \\
0 & \text{if } z \in \bar{m}
\end{cases}
\end{equation}

\STATE Compute feature vectors $\mathbf{x}(z) = [x_1(z), \dots, x_5(z)]$ for all $z \in \mathcal{D}$
\begin{equation}
x(z) = [x_1(z), x_2(z), x_3(z), x_4(z), x_5(z)] \in \mathcal{D}
\end{equation}

\STATE Train a logistic regression model:
\[
P(y = 1 \mid \mathbf{x}) = \sigma(\mathbf{w}^\top \mathbf{x} + b)
\]

\STATE Compute normalized weights:
\[
\alpha_i = \frac{|w_i|}{\sum_{j=1}^5 |w_j|}, \quad \forall i \in \{1, \dots, 5\}
\]

\STATE Compute per-sample membership score:
\[
m(z) = \sigma\!\left( \lambda \sum_{i=1}^5 \alpha_i x_i(z) \right)
\] where $\lambda > 0$ is a scaling parameter.

\STATE Compute the final benchmark score:
\[
\text{HALOB-MIA} = \frac{1}{|\mathcal{D}|} \sum_{z \in \mathcal{D}} m(z)
\]

\RETURN $\text{HALOB-MIA}$

\end{algorithmic}
\end{algorithm}

From Table \ref{tab:metric_data}, a representation of an AUC score of ~0.7 shows that a moderately strong attack with TPR/FPR 1$\%$ of 0.2 might be operationally useless in a real-world precision prevalence scenario. Suppose we have learned weights $\alpha$ = [0.35, 0.35, 0.20, 0.10, 0.10], an average HALOB-MIA score on members $\approx~0.55$, while non-members $\approx0.525$ which are near randomness.
This implies that traditional metrics such as AUC can mask negligible leakages, while a novel benchmark could significantly align the evaluation of MIAs with real-world leakage auditing constraints.

\paragraph{Traditional MIA Metric.}
\begin{equation}
\text{decision} = 1\big[\ell(z) < \tau\big]
\end{equation}

\paragraph{HALOB-MIA.}
\begin{equation}
\text{score} = \sigma\left( \sum_{i=1}^{5} \alpha_i \, x_i(z) \right)
\end{equation}

\section{Conclusion}
In this study, our main focus is on addressing why MIA models demonstrate good performance during evaluation but fail silently when they are deployed in real-world settings due to unconscious characteristics usually left unaddressed while training MIA models. We develop a tentative comparative analysis of four representative studies that deduce a number of important factors that affect the performance of MIAs in LLMs.

Over-reliance on traditional evaluation metrics for assessing MIA performance masks the vulnerability and privacy risks of the model, especially when the model performs well on random guessing. Post-hoc datasets bias also affects the performance of MIAs by introducing spurious data, which superficially increases the performance of attacks. Finally, we propose the development of holistic and standardized architecture-aware benchmarks that can accurately account for the performance of MIA models across all specified dimensions. We encourage subsequent work to prioritize multi-metric reporting and dataset evaluation pipelines that would prevent MIA models from failing post-deployment.

\bibliographystyle{unsrtnat} 
\nocite{*}\bibliography{references}


\appendix

\section{Appendix / Supplemental Material}

\begin{tcolorbox}[
    title=Key Position on MIA Evaluation Metrics,
    coltitle=black,
    colback=white,
    colframe=cyan,
    boxrule=0.5pt,
    left=5pt,
    top=5pt,
    bottom=5pt,
    breakable
]
\textbf{Insight:} MIA evaluations should adopt standardized architecture-aware benchmarks, multi-metric reporting and time-aligned sampling of
member and non-member sets that account for true MIA performance. 
\end{tcolorbox}

\begin{table*}[t]
\centering
\caption{Key observations from comparative analysis of MIA literature and their implications for future research.}
\label{tab:mia_observations}
\small
\begin{tabular}{|p{3.5cm}|p{4.5cm}|p{4.5cm}|}
\hline
\textbf{Observation} & \textbf{Evidence Across Papers} & \textbf{Implication} \\
\hline
Inflated metrics due to post-hoc sampling & 
PerProb~\cite{liao2025perprob}, Scaling Up MIAs~\cite{puerto2025scaling}, Limits of Strong MIAs~\cite{hayes2025exploring}  & 
Metrics may not reflect real memorization risk \\
\hline
Metrics fail at scale & 
Scaling Up MIAs~\cite{puerto2025scaling}, Do MIAs Work?~\cite{duan2024membership}& 
Small-scale attack success does not transfer to GPT-sized LLMs \\
\hline
Lack of standardized benchmarks & 
All papers & 
Hard to compare results or track progress \\
\hline
Architecture-sensitive attack behavior & 
PerProb~\cite{liao2025perprob} vs GPT studies~\cite{radford2018improving} & 
BERT metrics $\neq$ GPT metrics; cross-paper aggregation misleading \\
\hline
Over-reliance on single metrics & 
All papers & 
Accuracy or AUC alone is insufficient; multiple, context-aware metrics are needed \\
\hline
\end{tabular}
\end{table*}

\section{Perplexity as a Proxy for Memorization}
Liao et al.~\cite{liao2025perprob} uses perplexity differences as an indirect indicator of memorization. In a general context, MIA models with low training perplexity but high validation or higher test perplexity indicate that the model has memorized the training data set. They argue that such a lower perplexity model predicts sequences more confidently, which subtly implies that samples with lower perplexities indicate the model has memorized training data as compared to non-member samples. If the sequence is seen during model training, it is assigned a higher probability, implying a lower perplexity score. However, their efforts fail to capture memorization sequences alone. This also keeps count of distributional familiarity. For example, when a series of text samples is obtained in the same BERT-style domain, they are automatically assigned a lower perplexity, which contrasts with memorization. In such scenarios, the logical conclusion is that the model performs well in generalization on a familiar distribution.
For BERT-era models, they require less pretraining data compared to GPT-style models, which are trained on massive tokens. The perplexity difference could reflect memorization in BERT-style models to some extent, but it might fail extensively on GPT-style models~\cite{carlini2022quantifying}. Hence, the use of traditional metrics, such as perplexity, fails to account for true memorization, supporting our position for the adoption of a holistic benchmark to evaluate MIAs. In the analysis of MIA evaluations, memorization becomes extremely imperative because it directly interacts with the core dimensions, including attack type, dataset construction, metric results, and scaling effects.

\section{Results Reproducibility}
 Most studies have proven that MIAs in large-scale LLMs often perform near random guessing when the evaluation artifacts are controlled. Based on our analysis, such results are frequently excluded and evidently contribute to the skewed perception of the effectiveness of MIAs in LLMs and, as a result, hinder the progression in this field. It is imperative to fully understand when MIAs do not work and when they do. On a similar scale, it is vital to encourage negative and null results, and any reproduction studies should be treated as contributions in their own right.

\section{Limitations}
In this position paper, we do not explicitly propose new experimental evaluations for MIAs. While we argue that commonly used traditional MIA metrics fail silently at scale, we do not propose a single replacement metric, but instead, we advocate for a multi-metric approach centered around a standardized, holistic, and architecture-aware benchmark. Although this paper focuses on the risk of memorization as understood in a technical context, we do not pay additional attention to the legal context of privacy harm.

\noindent\textit{Conceptual Rather Than Empirical}: 
This paper mostly synthesizes and reinterprets previous results from the representative studies discussed.

\noindent\textit{Unified Alternative Metric Proposition}:  While we convincingly show the limitations of current metrics, we do not introduce a single alternative metric for the evaluation of MIAs.

\noindent\textit{Closed-Source and Industry Models}: This paper assumes a controlled evaluation environment, which might not be the case in the context of closed-source LLMs.

\begin{figure}[htbp]
    \centering
    \includegraphics[width=\columnwidth]{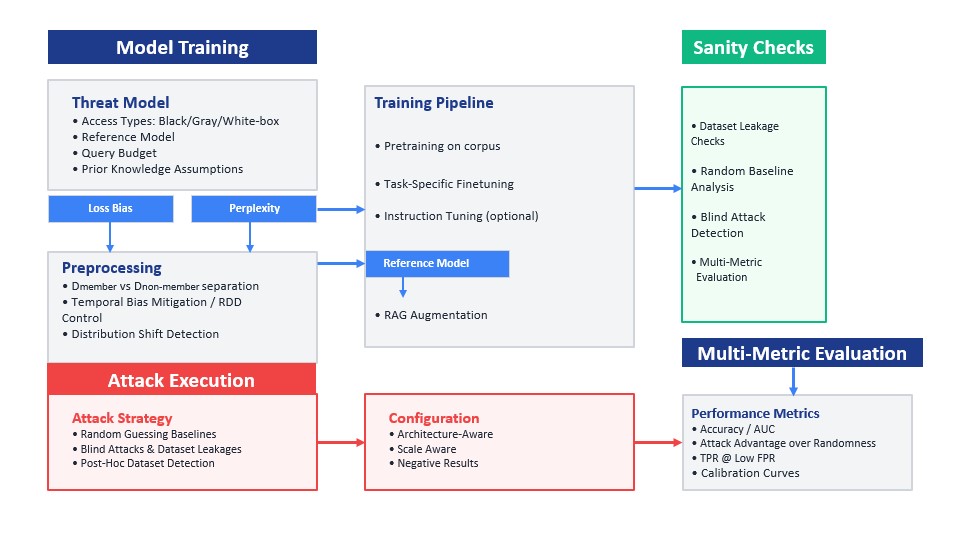}
    \caption{Membership Inference Attack Pipeline}
    \label{fig:attack_pipeline}
\end{figure}

\begin{table*}[t]
\centering
\caption{Core dimensions along which membership inference attacks (MIAs) on large language models differ, and why direct comparison across studies is often misleading.}
\label{tab:mia_dimensions}
\small
\begin{tabular}{|p{2.0cm}|p{4.5cm}|p{6.0cm}|}
\hline
\textbf{Dimension} & \textbf{Description} & \textbf{Analytical Rationale} \\
\hline
Attack Type &
Black-box vs. white-box; reference-model vs. direct attacks &
Different threat models assume varying levels of adversarial access, leading to fundamentally incomparable attack performance. \\
\hline
Model Architecture &
Encoder-only (e.g., BERT), decoder-only (e.g., GPT), instruction-tuned LLMs &
Architectural and training differences alter memorization patterns and leakage behavior, significantly affecting MIA success rates. \\
\hline
Dataset Construction &
Member/non-member splits; post-hoc sampling vs. curated datasets &
Dataset artifacts and leakage shortcuts can artificially inflate attack metrics, obscuring true memorization effects. \\
\hline
Metric Choice &
Accuracy, AUC, TPR, FPR, attack advantage, perplexity-based scores &
Different metrics capture distinct aspects of inference success; aggregate metrics may mask high false-positive rates or poor calibration. \\
\hline
Scale and Data Size &
Model parameter count, pretraining corpus size &
Scaling tends to reduce classical overfitting while introducing new forms of memorization, reshaping attack efficacy. \\
\hline
Baseline Selection &
Random guessing; dataset-only or threshold-based classifiers &
Baseline choice defines what constitutes ``success" and can exaggerate perceived vulnerability if poorly chosen. \\
\hline
Reporting Practices &
Confidence intervals, per-sample analysis, tail-risk metrics &
Reporting only peak or average performance can hide silent failure modes that emerge under deployment conditions. \\
\hline
\end{tabular}
\end{table*}

\section{Future Work}
To address the current limitations discussed on MIA evaluations, we intend to embark on critical steps towards performing a systematic re-evaluation of the existing MIA methods under controlled experimentation, with the inclusion of concentric initiatives to improve MIA evaluations.
 
\textit{Architecture-Aware Privacy Evaluation}: We intend to construct member and non-member sets that are within similar domain extracts in terms of source, period, pre-processing, and prompts, while documenting the implications of the various deviations.

\textit{Multi-Metric Benchmarking}: We intend to adopt a multi-metric benchmarking as opposed to the usage of a single-metric evaluation suite that might not be sufficient to evaluate the privacy risks.. 

\textit{LLM Specific Benchmarking Design}: We also look forward to working with LLM specific benchmarks instead of having to adapt classification-infused evaluation pipelines. This will facilitate a constant evaluation between studies and avoid any ambiguity associated with the results achieved. It will also be possible to factor in sequence length, prompt sensitivity, contextual effects, and different levels of granularity associated with memorization, such as token, span, document, and dataset-level granularity.

\textit{Analysis and Scaling of Negative and Null Results}: We also look forward to gathering, analyzing, and scaling results for different model sizes, data sets, and training paradigms to facilitate an understanding of when and where there is no leakage of privacy.

\textit{Simulating Silent Failures for Privacy Evaluation}: Lastly, we look forward to simulating deployed conditions and conducting stress tests for privacy evaluations to mimic how evaluation scenarios fail.

\end{document}